\PassOptionsToPackage{table,xcdraw}{xcolor}
\documentclass[manuscript, screen, sigconf]{acmart}

\usepackage{booktabs}
\usepackage{multirow}
\usepackage{subcaption}

\usepackage{tabularx}
\usepackage{adjustbox}
\usepackage{geometry} 
\usepackage{longtable} 
\usepackage{enumitem}
\usepackage{tcolorbox}
\tcbuselibrary{theorems}

\newtcbtheorem[number within=section]{hyp}{Hypothesis}%
{colback=grey!5,colframe=black!35!black,fonttitle=\bfseries}{th}

\AtBeginDocument{%
  }

\setcopyright{none}
\copyrightyear{\the\year{}}

\acmConference[Under Review]{Arxiv Preprint}{\today}{}
\acmBooktitle{Arxiv Preprint}

\usepackage{xspace}
\begin{document}
\newcommand{\LG}[0]{\textit{learning\_gains}\xspace}
\newcommand{\SI}[0]{\textit{score\_initial}\xspace}
\newcommand{\SF}[0]{\textit{score\_final}\xspace}
\newcommand{\WT}[0]{\textit{words\_typed}\xspace}
\newcommand{\CD}[0]{\textit{condition}\xspace}
\newcommand{\YR}[0]{\textit{year}\xspace}
\newcommand{\TU}[0]{\textit{usefulness}\xspace}
\newcommand{\TI}[0]{\textit{interestingness}\xspace}
\newcommand{\EC}[0]{\textit{comprehensiveness}\xspace}
\newcommand{\EI}[0]{\textit{level\_of\_resources}\xspace}

\title{GPT-4 as a Homework Tutor in Schools Can Improve Learning Outcomes and Engagement}
\title{GPT-4 as a Homework Tutor can Improve Student Engagement and Learning Outcomes}


\author{Alessandro Vanzo}
\authornote{Both authors contributed equally to this research.}
\email{alessandro.vanzo@inf.ethz.ch}
\affiliation{%
  \institution{ETH Zürich}
  \country{Switzerland}
}
\orcid{1234-5678-9012}

\author{Sankalan Pal Chowdhury}
\authornotemark[1]
\email{sankalan.palchowdhury@inf.ethz.ch}
\affiliation{%
  \institution{ETH Zürich}
  \country{Switzerland}
}
\orcid{0009-0005-8661-041X}

\author{Mrinmaya Sachan}
\email{mrinmaya.sachan@inf.ethz.ch}
\affiliation{%
  \institution{ETH Zürich}
  \country{Switzerland}}

\renewcommand{\shortauthors}{Vanzo et al.}

\begin{abstract}
 This work contributes to the scarce empirical literature on LLM-based interactive homework in real-world educational settings and offers a practical, scalable solution for improving homework in schools. Homework is an important part of education in schools across the world, but in order to maximize benefit, it needs to be accompanied with feedback and followup questions. We developed a prompting strategy that enables GPT-4 to conduct interactive homework sessions for high-school students learning English as a second language. Our strategy requires minimal efforts in content preparation, one of the key challenges of alternatives like home tutors or ITSs. We carried out a Randomized Controlled Trial (RCT) in four high-school classes, replacing traditional homework with GPT-4 homework sessions for the treatment group. We observed significant improvements in learning outcomes, specifically a greater gain in grammar, and student engagement. In addition, students reported high levels of satisfaction with the system and wanted to continue using it after the end of the RCT.
\end{abstract}




\maketitle

\section{Introduction}

Homework is an important component of education as it helps students self evaluate and develop self regulation skills  \cite{ramdass2011developing}. However, in order to fully benefit from solving homework problems, it is important that students receive swift feedback on their work  \cite{schartel2012giving}, which is not possible for teachers due to time constraints. This leads to several students having to resort to private tutors, which can be prohibitively expensive for several households  \cite{PrivateTutoringinItaly}. In this work we look at the possibility of 
leveraging GPT-4  \cite{openai2024GPT4technicalreport} as a tutor to assist students in their homework using a simple prompting strategy and an interface we developed. 

In the seminal paper on the 2 Sigma problem, Bloom  \cite{bloom2sigma} noted that compared to conventional learning, which consists of 
lectures followed by evaluation tests, students who received feedback on said tests and were given corrective instruction thereafter performed one standard deviation ($\sigma$) higher in terms of learning gains. The same paper notes a $0.8\sigma$ improvement for graded homework, but only a $0.3\sigma$ improvement if the homework is simply assigned without follow-up. The maximum benefit comes from one-on-one or one-on-few tutoring at $2\sigma$ but Bloom found no substitute for it. The status quo in schools roughly corresponds to the scenario where homework is never graded or, if graded, is done superficially. While the development of MOOCs with online lecture videos has extended the benefits of conventional education to larger populations \cite{10.1007/978-3-319-11200-8_8}, most MOOCs still lack proper homework feedback or corrective instruction. Attempts have been made to try to scale the benefits of feedback and corrective instruction 
with the 
use of peer-tutoring \cite{cohen1982educational} and Intelligent Tutoring Systems \cite{nwana1990intelligent} but these too have problems with effectiveness and scaling.

Recent developments in Large Language Models (LLMs)  \cite{openai2024GPT4technicalreport, geminiteam2024geminifamilyhighlycapable, touvron2023llamaopenefficientfoundation} have opened up the possibility of leveraging them for interactive homework and corrective feedback  \cite{KASNECI2023102274}. 
The rapid development and adoption of LLMs like GPT-4 have provided the educational community with several new opportunities as well as challenges. 
While the benefits of GPT models in education are well studied  \cite{baidoo2023education}, educators are also concerned by the potential use of GPT as a tool for plagiarism in homeworks  \cite{dehouche2021plagiarism} and them causing an overdependence by students  \cite{zhai2024effects}. 
These fears are further exacerbated by the prevalence of hallucinations  \cite{chelli2024hallucination} and jailbreaks  \cite{chu2024comprehensive}. 
Studies involving real-world situations are, therefore, extremely important.

Therefore, in this paper, we summarize our observations and learnings from testing out the effects of replacing static homework with a GPT-4 instance which has been instructed to cover the material of the homework, but in an interactive manner. 
Our intervention fills a gap that exists in many school systems without having to cause disruption to existing educational systems. 
We carry out a Randomized Control Trial (RCT) in an Italian high school to understand if such an intervention is indeed beneficial to students, both in terms of the students' self-assessments and also in terms of externally measured learning gains. 
We find that students who used GPT-4 (prepared as shown in Figure \ref{fig:teaser}) in this manner have limited improvements in learning gain, while also feeling better supported in terms of available resources. 
Finally, all students who would still be in school after the conclusion of the study indicated that they would want to continue to have access to the tutor, giving us hope that despite the contemporary fears that LLMs may lead to the decline of homework in education, LLMs can also be used to make homework more fun and didactically useful to students.





\section{Background}
\label{sec:otherwork}

Educational research has consistently shown that personalized tutoring is one of the most effective forms of instruction. However, scaling this approach has been a significant challenge. In recent years, advances in artificial intelligence, particularly through Large Language Models (LLMs) like GPT-4, have sparked interest in their potential to provide scalable, interactive tutoring solutions. Despite early successes, the use of LLMs in education has raised important questions around their efficacy, potential risks, and best practices for implementation.

In this section, we first explore the history and limitations of scaling tutoring using traditional methods and Intelligent Tutoring Systems (ITS). We then discuss the rise of LLMs and their application in education, focusing on their strengths and potential challenges. Finally, we review recent empirical studies that have evaluated LLM-based tutoring systems, highlighting both the promise and the gaps in the current literature, which our study aims to address.

\subsection{The Challenge of Scaling Tutoring in Education}

 In Bloom's seminal work on the "2 Sigma Problem"  \cite{bloom2sigma}, it was shown that students who received personalized tutoring performed two standard deviations better than those in traditional classroom settings. However, scaling one-on-one tutoring to large student populations is challenging due to the high cost and resource demands. Intelligent Tutoring Systems (ITS) attempt to fill this gap, promising to offer personalized, scalable instruction  \cite{nwana1990intelligent}. Yet, while ITS can help improve learning outcomes, they face challenges in terms of scalability, content preparation, and flexibility  \cite{cai2019authoring}.

\subsection{The Role of Large Language Models in Education}

The advent of Large Language Models (LLMs) such as GPT-4  \cite{openai2024GPT4technicalreport} represents a potential breakthrough in addressing the issue of scalability in tutoring. 
Unlike traditional ITSs, which require extensive manual content creation, LLMs can generate interactive and dynamic educational content with minimal human intervention. Research suggests that LLMs like GPT-4 can mimic tutor-like behavior by engaging in natural language dialogues, providing feedback, and offering corrective instruction, all of which are key components of effective tutoring  \cite{KASNECI2023102274}.

Despite their promise, the use of LLMs in education has sparked mixed reactions. On one hand, proponents argue that LLMs could offer an affordable and scalable tutoring solution, especially for students in under-resourced educational settings. 
Yet, many educators also express concern about issues like student over-reliance on AI tools, the risk of plagiarism, and the tendency of LLMs to hallucinate or provide incorrect information  \cite{dehouche2021plagiarism, zhai2024effects, chelli2024hallucination}.
Given these conflicting perspectives, empirical evidence from real-world classroom settings is critical to assess whether LLMs can truly enhance learning while mitigating potential risks.

\subsection{Empirical Studies on LLMs in Feedback and Interactive Exercises}

Several recent studies have begun exploring the effectiveness of LLMs and LLM-based systems, particularly in higher education and STEM fields \cite{padiyath2024insights, tanay2024exploratorystudyupperlevelcomputing, aruleba2023integratingchatgptcomputerscience,krupp2023unreflectedacceptanceinvestigating,odekeye4755024will}. A large volume of this work has focussed on LLM-based agents for programming support \cite{cref, qi2024knowledgecomponentbasedmethodologyevaluatingai, liffiton2023codehelpusinglargelanguage, Kazemitabaar_2024, Jacobs_2024, 10.1145/3626252.3630938, Lyu_2024, li2024tutorlyturningprogrammingvideos, choudhuri2023farwetriumphstrials, Pankiewicz_2024} However, most studies in this domain are limited to university-level computer science courses, and few offer rigorous control conditions for comparison.

Outside of computer science, a growing body of work has evaluated LLMs in domains like mathematics \cite{Chowdhury2024AutoTutorML, 10297824, 10.1371/journal.pone.0304013}, language learning \cite{doi:10.1080/2331186X.2024.2355385, Park_2024}, health sciences \cite{info:doi/10.2196/51344,chheang2024anatomyeducationgenerativeaibased, wang2024patientpsiusinglargelanguage} and other domains \cite{schmucker2024rufflerileyinsightsdesigningevaluating, thway2024battlingbotpoopusinggenai, chen2024effectiveness,zhang2024investigationeffectivenessapplyingchatgpt}. 
However, these studies often focus on highly structured tasks, such as answering specific questions or performing well-defined problem-solving activities, where the risks of misinformation are relatively low.

\subsection{Our Contributions}
Despite the numerous studies published in this field, we note that the literature on non-computer science subjects is still limited. In particular, empirical studies in primary and secondary schools are very few. We further note that there has been very little work involving school-aged \footnote{We refer to students younger than the typical age for tertiary education.} students and none of these gave their system the flexibility that we grant GPT-4 with our strategy.
The scarcity of studies does not necessarily imply a scarcity of potential for LLM-based technologies in this area. 
The potential of these models in schools is significant, largely untapped and well worth investigating.


Our work contributes to the understanding of the effectiveness and applicability of LLMs as tutors in schools. 
We contribute to the scarce literature on non-computer science subjects, and in particular to the even scarcer empirical literature. To the best of our knowledge, our work is the first in-field RCT incorporating a recent, state-of-the-art LLM as a tutor for language learning in a school.
Our design is: 
\begin{enumerate}
\item \textbf{Non-disruptive}: We intervene in the area of homework which has little involvement from the school system, via a prompting strategy needing no extra work from the teachers, thereby smoother adoption into the existing system
\item \textbf{Context Aware}: Our prompt informs GPT of what the teacher expects the student to cover in the current homework, maintaining the teacher's freedom to decide the curriculum contents and speed.
\item \textbf{Adaptable}: We can adapt to a wide variety of different homework, from simple fill-in-the-blank grammar exercises to short essays on complex topics.
\item \textbf{Minimalistic}: We limit the level of engineering to prompt design, allowing for the possibility of using more powerful LLMs in the future 
\end{enumerate}
We provide empirical insights into the potential of GPT-4 as a tutor which can be leveraged and built upon by the community in the future.

\section{Methodology}
\label{sec:methods}
\begin{figure*}[t]
  \includegraphics[width=\textwidth]{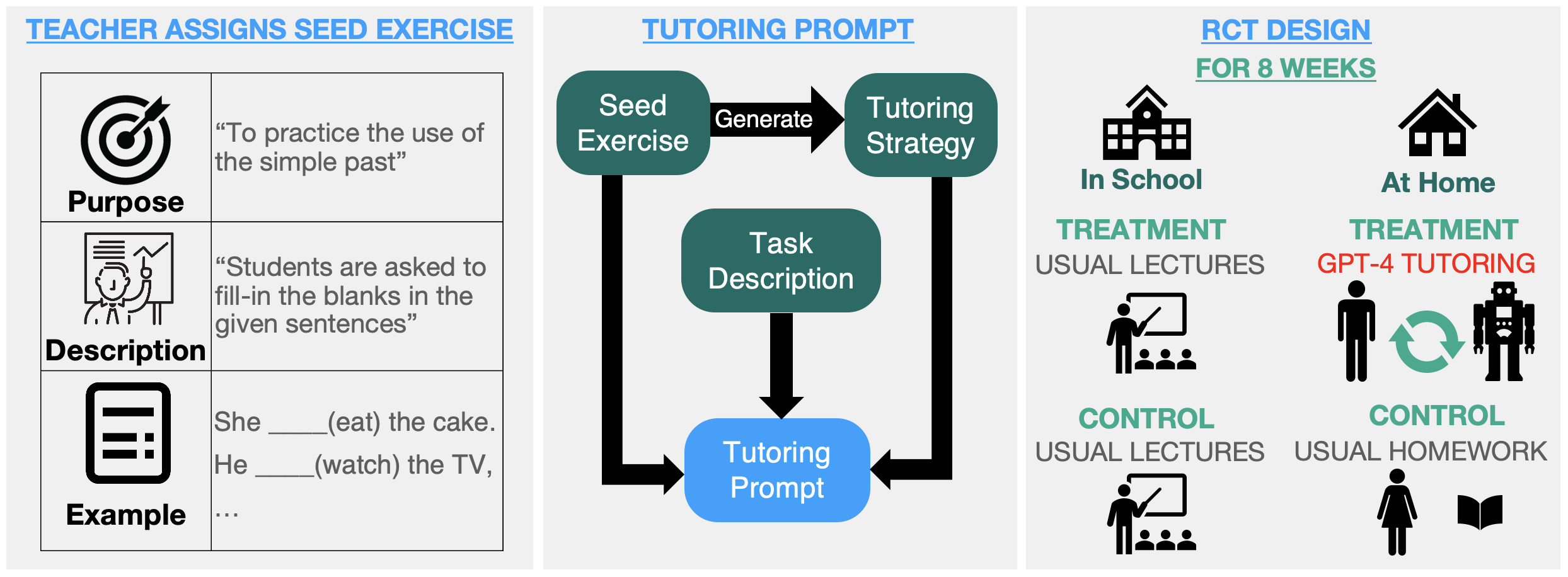}
  \caption{Illustration of the study design. We ask the teacher to provide the weekly homework exercises. For each exercise, we ask for three elements: \textit{purpose}, a brief, informal description of the learning goals; \textit{description}, outlining what the student is asked to do; \textit{example}, with an instance of the homework the student would typically be assigned. We prompt GPT-4 with a description of the tutoring task and with the 3 elements of the exercise, asking to cover the same concepts and pedagogical purpose. Finally, we test the effectiveness of the tutoring as a replacement for standard homework in an RCT.}
  \Description{}
  \label{fig:teaser}
\end{figure*}

We provide an overview of the study design in figure \ref{fig:teaser}. The teacher, who assigns weekly homework exercises to students, provides three key elements for each exercise: the \textit{purpose}, a brief informal description of the learning objectives; the \textit{description}, outlining the specific tasks students are asked to complete; and an \textit{example}, representing a typical instance of the homework assignment. We prompt GPT-4 with these elements, instructing it to generate an interactive exercise session aligned with the original pedagogical goals. We test the effectiveness of GPT-4 as a tutoring tool compared to traditional homework in a randomized controlled trial (RCT). We assess student's learning outcomes and engagement, using both external measures (pre- and post-tests) and student feedback through questionnaires. 

In this section, we first briefly describe the background of the participants and the area of intervention. We then describe the prompting strategy. Finally, we describe the RCT design and the questionnaires that were used.

\subsection{Participants}
We conducted our study in a high school in Italy.
The Italian high school system, “scuola superiore,” spans 5 years and includes lyceums, technical institutes, and professional institutes. Lyceums prepare students for university, technical institutes offer career-oriented education, and professional institutes provide vocational training.

We partnered with a technical institute, working with 4 classes, all of whom were taught English by the same teacher. Two classes were in the 3rd year (median student age 16) and two in the 5th year (median age 18) on the tourism track, focusing on business administration and foreign languages. The 3rd year classes consisted of a total of $39$ students, of which $20$ ($18$F and $2$M) were assigned the control group while $19$ ($17$F and $2$M) were assigned the treatment group. The 5th year consisted of 37 students, of which $19$ ($17$F and $2$M) were assigned to the control group and $18$ ($13$F and $5$M) were assigned to the treatment group. All the students had access to internet either through a smartphone or through a computer.

\subsection{Area of intervention}
For every class, the English curriculum was composed of two main parts: 3 hours of weekly lectures and 1-2 hours of weekly homework and self-study. 
We intervened on homework and self-study. Treatment group students were assigned interactive sessions with GPT-4. Control group students were assigned the typical homework they would have for the rest of the year.
Treatment and control group students attended the same lectures. An overview of the homework assigned in each class can be found in table \ref{tab:homework}.


It must be noted that the students of neither group were forbidden from using ChatGPT on their own\footnote{Almost two-thirds of Italian students are likely to be using ChatGPT according to \href{https://tg24.sky.it/tecnologia/2024/05/20/intelligenza-artificiale-scuola-chat-gpt}{this} report} so any effects seen are in addition to that of self-usage.



\subsection{Prompting Strategy}
\label{prompting_strategy}
Figure \ref{fig:teaser} left and centre panel summarize our prompting strategy. Our goal is to give GPT4 sufficient information in order to align the homework with the expectations of the teacher, without requiring any LLM-related expertise from the teacher or imposing any additional workload on them. As such, we ask the teacher to provide the following 3 components of the \textit{seed exercise}: 

\begin{enumerate}
    \item \textbf{Exercise purpose:} the pedagogical purpose of the exercise, described in a few sentences 
    \item \textbf{Exercise description:} a description of the task 
    \item \textbf{Exercise example:} the exercise itself as it would be shown to the students 
\end{enumerate}
We note that the teacher would need to prepare these for regular homework anyway, and the only additional work is entering these to the platform instead of disbursing it to the students. In post-study feedback, the teacher claimed that while this change introduced an initial learning curve for them, in the long term it would reduce their workload somewhat.

Having obtained the seed exercise, we first ask GPT to generate a step-by-step plan on how to carry out the homework. The final prompt is obtained by appending the seed exercise and the generated strategy to a generic Task Description. The LLM of our choice was \texttt{gpt-4-0125-preview}. The interface was hosted on a dedicated website that students could access with their own personal devices outside of school hours. 

We report the GPT-4 prompts for strategy generation and tutoring in the appendix \ref{app:prompt}. For the strategy generation, we provide the seed exercise alongside a basic description of the tutoring task, mentioning that we are working with high school students and aiming at a B2 level of English according to the Common European Framework of Reference for Languages (CEFR), and ask for an appropriate tutoring strategy. In the tutoring prompt, we provide a description of the tutoring task together with the seed exercise and the generated strategy.

\subsection{Randomized Controlled Trial}
\subsubsection{Research Goals}

Our Primary focus is on how much the tool helped (or hindered) student learning, and how the students felt about using the tool both in the short and the long term. Beyond this, we also look at their general outlook to the tool, how much they used it, and which groups benefited more. We use a combination of external measures(tests) and internal measures(questionnaires) to achieve this, as detailed in the rest of this section.

\subsubsection{Intervention Design}
 
We assigned students within each class to either the treatment or the control condition using stratified randomization based on their self-reported English GPA in the current year.
The teacher was not informed of the condition of the individual students, and lectures were the same for all students. The students were instructed not to share their group with her, to avoid interference.

All students received weekly homework consisting of one or more exercises, and the homework was carried out on a dedicated website that students could access with their own devices, including smartphones and computers.
Students in the control group received the homework as assigned by the teacher, in a format comparable to the exercise they received outside of the experiment. They solved each exercise individually and uploaded the answer on the online platform. For each exercise assigned by the teacher, students in the treatment group had access to a chat with GPT4. In each chat, the tutor used the prompt for the specific exercise. Students did not see the original exercise from the teacher. In both treatment and control conditions, students could access the platform at any time. We did not enforce a minimum or maximum level of engagement. The teacher could not observe the content of the student's solution, but we informed her of who had submitted a solution for each exercise. The intervention was planned to run for 6 weeks, and was extended to 8 weeks due to delays in covering the required content.

\subsubsection{Questionnaire design}
Students completed one questionnaire before the beginning of the intervention, one questionnaire per week during the intervention, and one questionnaire after the end of the intervention. We refer to these questionnaires as initial questionnaire, weekly questionnaire and final questionnaire, respectively. In addition, students had to complete a pre-test and post-test on the content covered during the intervention.\\

\noindent\textbf{Initial Questionnaire:} This questionnaire was given to the students before the beginning of the experimental intervention. It asked them to provide some contact information and basic information about themselves (name, age, email). 
We then included 9 background questions about the student performance and motivation at school, some general and some English-specific. We did not find a suitable standardized questionnaire, so we developed these questions specifically for our experiment.
Afterwards, we included standardized questions on self-efficacy at school. We selected the questions from \cite{selfEfficacy}, translated them to Italian and applied some minor adjustments to make them specific to English. This subsection included 6 questions.
 We then included 6 questions about the student experience with English homework, each question targeting one of the 4 ARCS aspects \cite{arcs}. While the ARCS model was initially developed to guide the development of education content, it is not uncommon to see questionnaires targeting its 4 main aspects to assess the effectiveness of educational contents.
Finally, we included 6 questions assessing the student experience during lectures, also targeting the ARCS dimensions.


\noindent\textbf{Final Questionnaire:}
The final questionnaire largely mirrored the initial questionnaire. However, instead of asking about the experience of students at school in general, it asked specifically about the experience over the time the experimental intervention ran (2 months).
In addition, we included questions for the treatment group students asking feedback about about the tutor.

\noindent\textbf{Weekly Questionnaire:}
In the weekly questionnaire, we included 3 questions about the exercise session as a whole and 2 questions for each exercise, using a 6-point Likert scale. In addition, for the treatment group, we asked to report how the tutor was helpful giving the option to select among a range of potential useful aspects.


\noindent\textbf{Pre-Test and Post-Test:}
Both the pre-test and post-test consisted of 24 multiple-choice questions. We assigned 1 point for each question answered correctly. The questions were provided to us by the teacher, who designed 8 questions for each week of the intervention. For each week, we randomized half of the questions to the pre-test and half to the post-test.
\newline\newline
\noindent We report all questionnaires in Appendix, table \ref{tab:app_final_q}.

\section{Results and Analysis}
\label{sec:results}


We start off with some general statistics, and then proceed to discuss each of our research questions in their own section.
\begin{table}[ht]
\centering
\resizebox{0.98\linewidth}{!}{
\begin{tabular}{@{}lc|c|c@{}}
\toprule
\textbf{} & \textbf{Third Year} & \textbf{Fifth Year} & \textbf{Total} \\ \midrule
\textbf{Control Group} \\\midrule
\# Assignments & 195 & 97 & 292 \\
Median Homework Word Count & 56 & 157 & 74 \\
\midrule
\textbf{Treatment Group} \\\midrule
\# Chats & 199 & 93 & 292 \\
Agent Messages (Median)& 15 & 12 & 14 \\
User Messages (Median)& 14 & 11 & 13 \\
Total Words per Chat - Agent (Median)& 977 & 1040 & 989 \\
Total Words per Chat - User (Median) & 114 & 314 & 143 \\ 
\bottomrule
\end{tabular}}
\caption{Usage statistics for the tutor}
\label{tab:cont}\label{tab:usage_summary}
\end{table}

\subsection{General statistics}
\subsubsection{Participant Background}
Table \ref{tab:grad} reports the self reported participant backgrounds. 
We note that there is a slight overestimation of english ability on part of the students, as more people consider themselves above average than below average as can be seen in table \ref{tab:grad}
\subsubsection{Usage Statistics}
\begin{table}[]
\centering
\resizebox{0.98\linewidth}{!}{
\begin{tabular}{@{}lc|c|c@{}}
\toprule
\textbf{} & \textbf{Third Year} & \textbf{Fifth Year} & \textbf{Total} \\ \midrule
\textbf{Control Group} \\\midrule
\#Students & 20 & 19 & 39 \\
Mean Grade in English & 7.20 & 7.54 & 7.37 \\
Held Back in English & 1 & 3 & 4 \\
Below Average English Ability &4&7&11\\
Average English Ability &8&3&11\\
Above Average English Ability &8&9&17\\
\midrule
\textbf{Treatment Group} \\\midrule
\#Students & 19 & 18 & 36 \\
Mean Grade in English & 7.63 & 7.43 & 7.53 \\
Held Back in English & 0 & 1 & 1 \\
Below Average English Ability &6&6&12\\
Average English Ability &5&5&10\\
Above Average English Ability &8&7&15\\
\bottomrule
\end{tabular}}
\caption{Self reported previous performance by students}
\label{tab:grad}
\end{table}
Table \ref{tab:cont} shows the overall usage summary of the platform. The 5th year homework consisted of open questions on literature and history, with several questions each week. The most common behaviour among students was to write a somewhat complete answer. The answer was then refined iteratively based on feedback from the tutor, adding nuance, correcting grammar and including or fixing factual information.
The 3rd year homework consisted of objective type (except for one essay type exercise), where the students were given sentences which they had to edit, complete or transform according to the question, and there was almost always a single correct answer. Student utterances for these questions were most of the time just attempts at the right answers, and not many students tried to have full conversations. We segmented the conversations into a total of $1549$ questions, of which $940$\footnotemark \, were solved immediately by the students, while in $365$\footnotemark[\value{footnote}]  cases the tutor revealed the answers. The conversations where a reveal occurred were on average $4.7$ utterances long, which would imply about $2$ attempts by the student. Correct cases were almost always $3$ utterances long (Tutor-Student-Tutor) with the exception of an exercise that required both the passive form and the double object passive form which required $5$ utterances.
\footnotetext{As judged by GPT-4o. These might have some errors}
\subsubsection{General Outlook}
As a part of the final survey, students in the treatment group were asked how they felt about the tutor. 32/33 respondents thought that the tutor helped them with their homework, whereas 30/35 felt that the tutor improved their on a practical level. Further, 26/34 respondents felt that the tutor helped them keep up with the English program.  Most importantly, 32/35 overall respondents wanted to continue using the tutor, with the 3 people saying "No" all being in their last year of High School. This shows that even in its current basic form, students enjoyed the experience of using it.
\subsection{Primary Analysis}
In this section we present the primary observations of our study. All variables(mentioned in \textit{italics} map to questions in one of our questionnaires. for more details, see section \ref{questionnaires} in the appendix.
\subsubsection{Overall Learning Gains}
We start off by comparing overall learning gains of students in the two conditions. To evaluate this, we conduct a one-sided t-test and obtained a Cohen's d of $0.251$ with a p-value of $0.314$. However, since the curricula for the 5th and 3rd years are substantially different, resulting in a significant disparity in score improvements between the two cohorts ($d=1.347,\,P<0.001$), we perform distinct tests for each class. For 3rd year the effect side is much larger favouring the treatment group, and is marginally significant ($d=0.603,\,P=0.087$). For the 5th year, there is almost no difference ($d=-0.004,\,P=0.991$).  We posit that this difference could emerge due to the 5th year homework being essay type compared to the third year homework being more objective with a single correct answer, which could have led to the following $2$ issues:
\begin{enumerate}
    \item The lack of a clear correct answer would make 5th year answers harder to evaluate.
    \item The pre- and post-test for both classes was objective type so the 5th year homework would have helped less in general.
\end{enumerate}
Overall, given the limited nature of intervention, we can conclude that under the right conditions, the treatment group does perform better.
\subsubsection{Short Term Experience of Students} We had weekly questionnaires for the entire week and also at exercise level. The week level questionnaires asked students about the \TI and \TU if their homework (see table \ref{tab:app_weekly_q} for full texts) while the exercise level questionnaires asked about the \EC and \EI of that particular exercise (see table \ref{tab:app_weekly_q} for full texts). We observed that students in the treatment group gave higher ratings in all 4 categories. Of these \TI($d=0.593,\, P=0.011$) and \EI($d=0.586,\, P=0.015$) were significant but \TU($d=0.356,\,P=0.125$) and \EC($d=0.281,\,P=0.234$) were not significant. Further, the treatment group were asked what they liked about the tutor, and 93\% of the times, the students picked atleast one option. The overall responses to this question are summarised in Table \ref{tab:like}.

\begin{table}[ht]
\centering
\resizebox{0.98\linewidth}{!}{
\begin{tabular}{l r}
\toprule
\textbf{Useful Aspect} & \textbf{Proportion (\%)} \\
\midrule
The tutor's explanations                             & 63\% \\
Receiving feedback and corrections on my answers     & 57\% \\
Being guided step by step in the solution            & 45\% \\
Solving the exercise itself                          & 25\% \\
Other                                                &  0.42\% \\
\midrule
\textbf{Summary} & \\
\midrule
Percentage Marked At Least One                       & 93\% \\
Percentage Marked 2 Useful Aspects                   & 66\% \\
Percentage Marked 3 Useful Aspects                   & 29\% \\
Percentage Marked 4 or More Useful Aspects           &  3.8\% \\
\bottomrule
\end{tabular}}
\caption{Student responses to what they liked about each exercise, treatment group only}
\label{tab:like}
\end{table}

Overall, given the responses, we conclude that most of the students found at least one facet of the tutor useful, making their overall outlook positive.
\subsubsection{Long Term Experience of Students}
\begin{figure*}[h]
  \centering
  \includegraphics[width=1.0\linewidth]{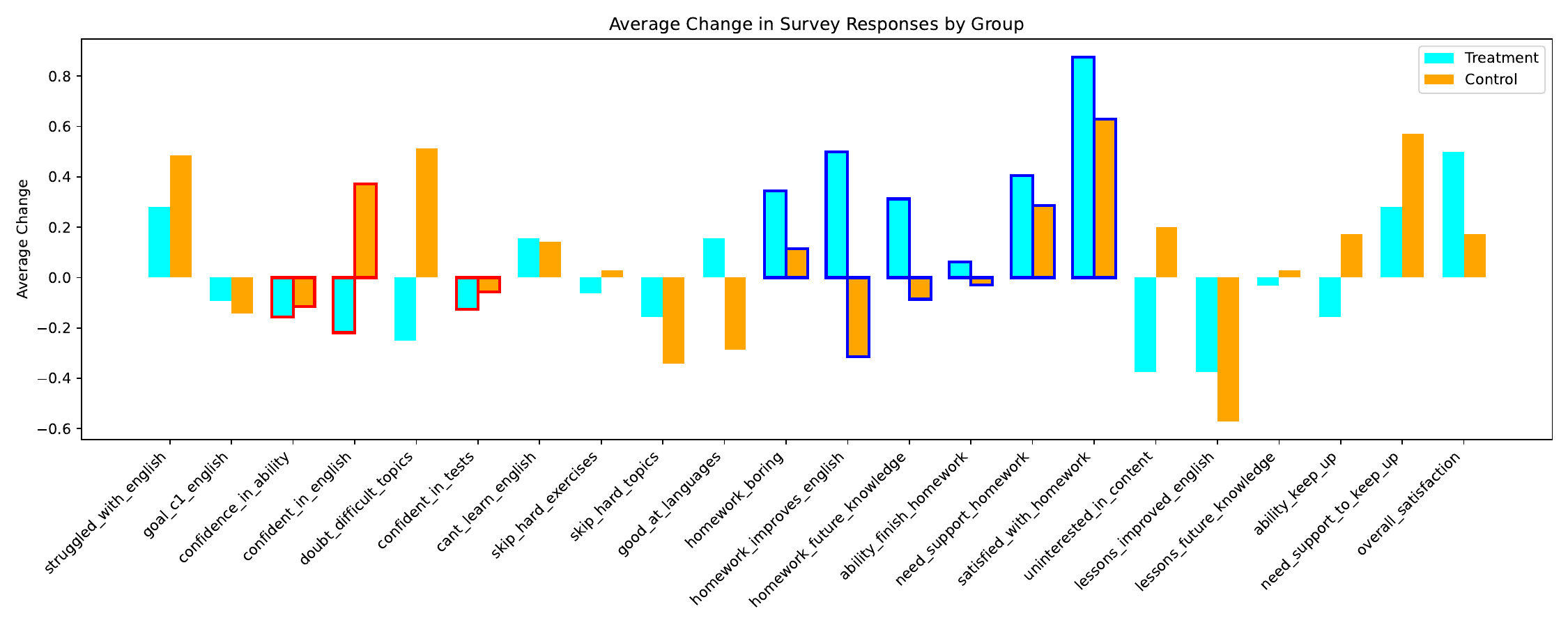}
  \caption{Change in Survey responses between the initial and the final questionnaire. Questions with negative sentiment are flipped(these are marked with a $\dagger$ in tables \ref{tab:app_initial_q} and \ref{tab:app_final_q}) so higher is always an improvement. Questions regarding confidence are marked in \textcolor{red}{red borders}. Questions regarding homework are marked in \textcolor{blue}{blue borders}. We notice that treatment group does better than control group on all questions regarding homework, which is a good sign for GPT4. We also note that they do worse in questions about ability}
  \label{fig:average_change_plot}
\end{figure*}

Both the initial and final questionnaires included 22 questions based on SESQ and ARCS frameworks (see tables \ref{tab:app_initial_q} and \ref{tab:app_final_q} for full texts of the questions). Figure \ref{fig:average_change_plot} shows the average change in the students' responses to these between the initial and final surveys (questions where a higher rating would be more negative have had their signs reversed to make higher is better for all questions). We note that note that the differences in the two groups are not significant for any of the questions (after correcting for FDR) but still certain trends emerge. First of all, overall satisfaction increases for both groups, but increases more for the treatment group. Further, for all questions relating to homework (where we intervened), the treatment group's opinions improved more than that of the control group. 

Another interesting trend that we notice is that students' confidence actually decreased for the treatment group. One possible reason might be that the students were mildly overconfident in their ability to begin with, as is indicated by the self-assessment of ability in English. However, since we did not reassess ability in the final questionnaire, this is hard to verify.
\subsection{Secondary Analysis}
In this section we investigate test our data for the existence of effects observed in previous research.
\subsubsection{Do stronger students benefit more?}
Next, we look at the relation between initial skills of student and how much they benefit. This is interesting because previous work \cite{prather2024wideninggapbenefitsharms,cipriano2024chatgpthelpreplaceanybody} suggests that students with higher initial knowledge benefit more from GPT Tutoring. In order to evaluate this, we calculated the Pearson-R and found that there was in fact a negative correlation ($R=-0.777,\,P<0.001$) between \SI and \LG which actually stronger than the control group($R=-0.628,\,P<0.001$). This indicates that weaker students actually benefit more from the tutoring compared to stronger ones. The trend holds individually for the treatment groups of the third year ($R=-0.667,\,P<0.005$) and the fifth year ($R=-0.843,\,P<0.001$). Although it is unlikely that ceiling effects influenced this, as the highest overall score was 2 points below the maximum possible $24$, the difference could be due to the fact that the mapping between knowledge and score is not linear and it is harder to improve a score that is already good. The difference from previous literature might be due to difference in domains, instruments, participants etc. and probably needs to be explored in future work.
\subsubsection{Effect of Engagement on Learning gains}
 \WT and \LG were positively correlated ($R=0.316,\, P=0.009$). The correlation is largely driven by the 3rd year ($R=0.434,\, P=0.007$). This correlation is stronger in the 3rd year treatment group ($R=0.454,\, P=0.077$) than the 3rd year control group ($R=0.264,\,P=0.275$). Similar trends are observed for the 5th year, although all correlations are non-significant. We further note that, overall, the \WT is significantly higher for the treatment group ($d=1.421,\,P<0.001$) and on running an OLS Regression  for  \LG wrt \CD, \WT and \YR, we find that the coefficient for the treatment condition is non-significant ($coef=-0.446,\,P=0.666$)\footnote{see Table \ref{tab:eng} for all coefficients}. This could mean that the benefit of the treatment condition is largely mediated by engagement, which is consistent with previous work \cite{altememy2023ai,doi:10.1080/23311916.2024.2353494}

\subsubsection{Hallucination and Other Errors}One major concern to deployment of GPT into educational scenarios is the fact that it can give incorrect information, which can in turn harm learning in students. To test the level of such errors, students were asked every week if the tutor had made some errors in their chats. Of 160 responses, only $16$ indicated that there was a problem. Only one of these was from year $5$, about the exercise being too long which is not a hallucination. Of the remaining 15, 10 instances referred to single point errors, i.e., at most $1$ sentence had an issue. A further $2$ responses complained on the nature of the exercise, which is again no a hallucination. Only $3$ responses said that multiple sentences were wrong ($3$ of which came from the same user)/ We went over all the conversations from the given weeks, and found a total of $4$ errors: In one case the tutor suggested a change of verb, but then suggested not changing it when the student agreed, another where the correct answer was not in the options, one where the tutor rejected a correct answer that was different from what it expected and one where both options provided were correct depending on context. Overall, this means that the tutor made no more than $14$ errors over $1549$ questions (and never doubled down on its errors) giving a hallucination rate of less than $1\%$ which is well within acceptable thresholds.

\subsection{Novelty Effects}
To investigate the novelty effects we look at whether students judgement of \EC, \EI, \TU and \TI over the weeks. Since the homeworks as well as weekly surveys are entirely voluntary, not every student filled in the surveys for each week\footnote{the surveys were designed to be instituted after finishing lessons according to the original lesson plan, so despite the actual experiment running for 8 weeks, we did only 6 surveys}. The presence of a novelty effect would be indicated by a drop in ratings over the weeks. To test this, we plot box plots of all the ratings given by the students.
\begin{figure}[h]
  \centering
  \begin{subfigure}[h]{\linewidth}
  \includegraphics[width=\linewidth]{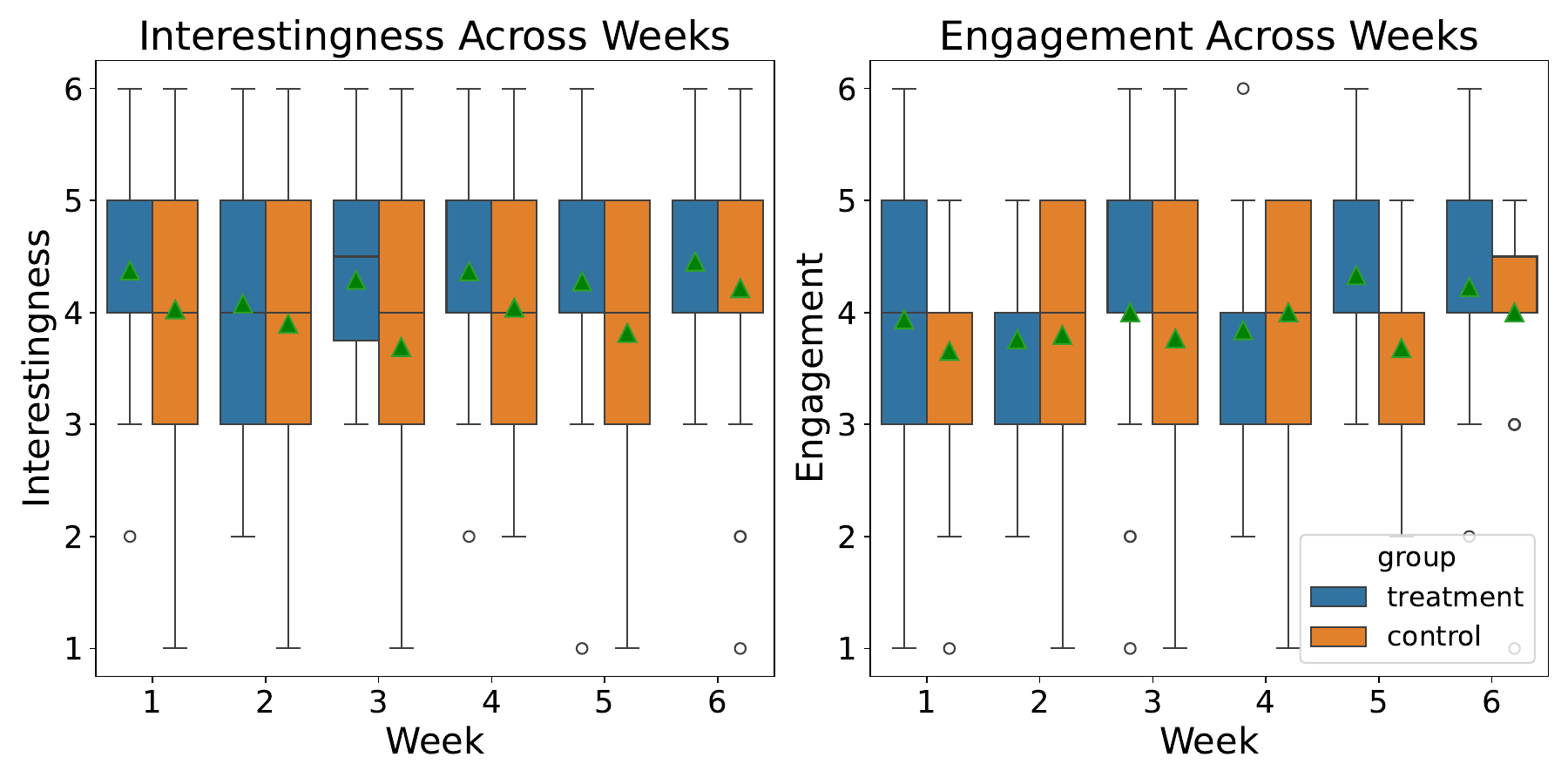}
  
  \end{subfigure}
  \begin{subfigure}[h]{\linewidth}
  \includegraphics[width=\linewidth]{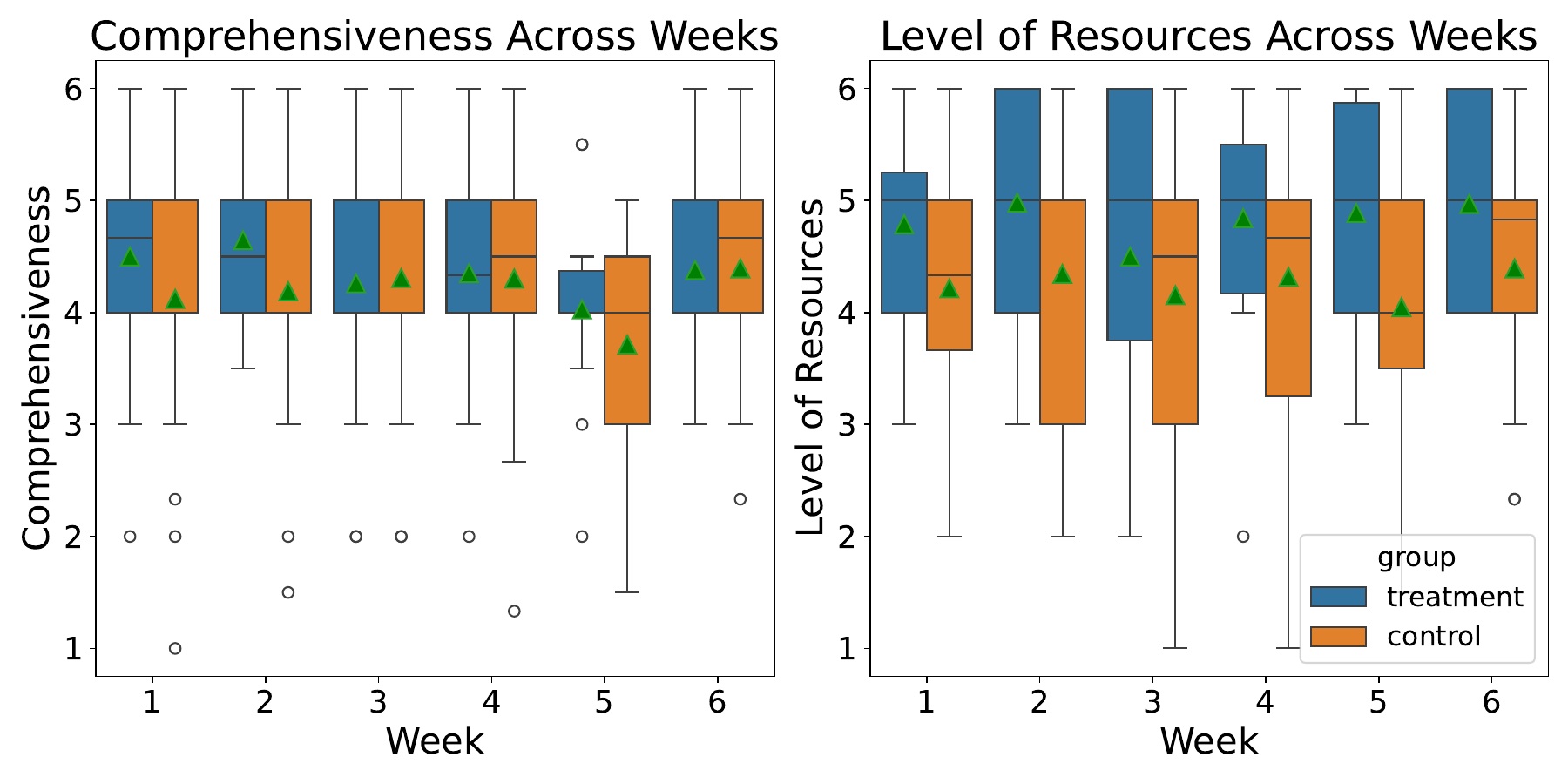}
  \end{subfigure}
  
  \vspace{-4mm}
  
  \caption{Weekly distribution of student ratings for \TU, \TI, \EC and \EI. \textcolor{green}{Green} triangles show means We observe no significant decline in any measure over time so there is no support for the existence of novelty effects}\
  \label{fig:novelty_w}
\end{figure}

The results of the tests are shown in Figure \ref{fig:novelty_w}.
For \TI, \EI and \EC, all the quartiles overlap, with the means lying in the region of overlap. \TU seems to vary from week to week, but the pattern is quite random, and no steady decline is visible. While this is not sufficient to show a lack of novelty effect, we can still say that it is not strong enough to pose a threat to the validity of our study.
\section{Conclusions and Discussion}
\label{sec:discussion}
In this work, we run an RCT to evaluate the ability of GPT-4 to function as a tutor. We find that students find this replacement of homework more useful and interesting, and are enthusiastic about continuing using it in their education. In addition to this, we also observe some improvement in learning as measured by tests. Also, we do not find evidence of bias towards stronger students or harmful hallucinations. We further notice that the self-assessments don't show significant decline over the RCT period, thereby making novelty effects less likely.

We do observe some issues with the tutor revealing answers too much, especially when students try to game the system, but the benefits seem to outweigh the drawbacks. The learning gains are also much smaller than the potential 2-sigma improvement, but this can be attributed to the small time scale. Further work in this field could explore
\begin{itemize}
    \item Extension of the tutor to other languages and subjects
    \item More directed prompts for specific types of exercises
    \item Including aspects of student modelling and pedagogical strategies
    \item More long-term effects of  AI-based tutoring
\end{itemize}

School and homework are often perceived as an unwelcome chore by students, and according to the personal experience of some teachers we worked with, there is an increasing lack of interest and engagement from the students, making interventions on these aspects even more needed. In addition, students more often felt that they had the necessary resources to complete what was required for them. Many parents cannot afford after school services and personal tutoring for their children, and struggling students are often left lagging behind. Empowering a larger number of students with the resources needed to achieve what is expected for them has the potential to reduce the gap between students living in more and less privileged circumstances, providing a fairer playing field within schools.
Given the continuing development of LLMs to improve them across all tasks, we believe that our study shows us the glimpse of a very bright future where hard to scale tasks like tutoring can be taken over by AI tutors, bringing the benefit of tutoring to a much greater number of students around the world.
\section{Ethics Statement}
This study was approved by the Human Subjects Committee of the Faculty of Economics, Business Administration, and Information Technology of the University of Zurich (OEC IRB \# 2024-018).
All participants provided informed consent prior to enrollment. For minors, the consent was provided by both parents or legal tutors, unless one parent or tutor alone had sole tutorship. Participant confidentiality was strictly maintained, and all data were anonymized. The study complied with all applicable regulations and ethical standards. Each aspect of the design was discussed and developed in collaboration with education professionals and with the school personnel involved, to ensure an optimal and fair experience for all participants and non-participants. The school personnel was informed of the potential risk and instructed to carefully monitor participants as well as non-participating students potentially affected by the study, and to provide them with extensive support.

\begin{acks}
We extend our sincerest thanks to the teachers, staff, and administration of Istituto Pindemonte in Verona for their support in conducting this study, especially Dr. Lara Quarti, in whose classes the experiment was conducted. We thank Dr. Julia Chatain for her guidance on the framing of this study. We thank Shashank Sonkar, Shehzaad Dhuliawala and Manuel Bernal-Lecina for their feedback on the final drafts of the paper.

Sankalan Pal Chowdhury additionally thanks his colleagues Manuela Pineros-Rodriguez, Fan Wang and Adrienn Toth for their helpful insights during discussions about this project. He is partially funded by the ETH-EPFL Joint Doctoral Program for Learning Sciences.

This project was funded by a grant from the Swiss National Science Foundation (Project No. 197155) and a Responsible AI grant by the Haslerstiftung.
\end{acks}

\bibliographystyle{ACM-Reference-Format}
\bibliography{sample-base}

\onecolumn
\appendix


\section{Questionnaires}
We report the initial questionnaire, the final questionnaire, and the weekly questionnaire. We report the orginal questions in Italian as well as an English translation.
\label{questionnaires}

\begin{scriptsize}
\begin{longtable}{@{}p{0.25\linewidth}p{0.25\linewidth}p{0.2\linewidth}p{0.2\linewidth}@{}}
\caption{Initial Questionnaire. Questions with negative sentiment are marked with a dagger(\textsuperscript{\dag})}
\label{tab:app_initial_q}
\\
\toprule
\textbf{Original Question} & \textbf{Translation} & \textbf{Type of Answer} & \textbf{Dataset Label} \\
\midrule
\endfirsthead
\multicolumn{4}{c}%
{\tablename\ \thetable\ -- \textit{Continued from previous page}} \\
\toprule
\textbf{Question} & \textbf{Translation (English)} & \textbf{Type of Answer} & \textbf{Label} \\
\midrule
\endhead
\midrule \multicolumn{4}{r}{\textit{Continued on next page}} \\ \midrule
\endfoot
\bottomrule
\endlastfoot

Età & Age &  & \texttt{age} \\ \midrule
Sei mai stato rimandato in inglese? & Have you ever been held back in English? & Yes; No & \texttt{failed\_english} \\ \midrule
Come pensi che sia il tuo livello di inglese rispetto alla media della tua classe, al di là dei voti? & How do you think your level of English compares to the average in your class, aside from grades? & Below average; Slightly below average; Average; Slightly Above Average; Above Average & \texttt{english\_level} \\ \midrule
Qual'è la tua media in inglese quest'anno? & What is your average grade in English this year? & $<$6; 6-6.49; 6.5-6.99; 7-7.49; 7.5 - 7.99; 8 - 8.49; 8.5-9; $>$9; I would rather not disclose & \texttt{english\_average} \\ \midrule
\multicolumn{4}{c}{\textbf{General Questions}: Questions marked with a `*' are standardized SESQ.} \\

\midrule
Sono molto motivato/a a studiare per raggiungere buoni risultati a scuola. & I am very motivated to study to achieve good results at school. & 6-point Likert & \texttt{motivated\_to\_study} \\ \midrule
Mi impegno molto per avere buoni risultati in inglese. & I work hard to get good results in English. & 6-point Likert & \texttt{effort\_in\_english} \\ \midrule
Faccio fatica a stare al passo col programma di inglese & I struggle to keep up with the English program & 6-point Likert & \texttt{struggled\_with\_english\textsuperscript{\dag}} \\ \midrule
Voglio raggiungere un buon livello di inglese nei prossimi anni & I want to achieve a good level of English in the coming years & 6-point Likert & \texttt{goal\_c1\_english} \\ \midrule
Penso di avere le capacità per raggiungere un buon livello di inglese nel corso dei prossimi anni & I think I have the capacity to reach a good level of English in the coming years & 6-point Likert & \texttt{confidence\_in\_ability} \\ \midrule
Avrei bisogno di più supporto e risorse per riuscire a raggiungere un buon livello di inglese & I would need more support and resources to achieve a good level of English & 6-point Likert & \texttt{need\_more\_support} \\ \midrule 
* In inglese, indipendentemente da quanto un argomento è difficile, sono sicuro di poterlo capire. & In English, no matter how difficult a topic is, I am sure I can understand it. & 6-point Likert & \texttt{confident\_in\_english} \\ \midrule
* Non sono sicuro di poter imparare gli argomenti più difficili del programma di inglese. & I am not sure I can learn the most difficult topics of the English program. & 6-point Likert & \texttt{doubt\_difficult\_topics\textsuperscript{\dag}} \\ \midrule
* Sono sicuro di poter andare bene in verifiche, interrogazioni e/o test di inglese. & I am sure I can do well in quizzes, oral exams, and/or English tests. & 6-point Likert & \texttt{confident\_in\_tests} \\ \midrule
* Per quanto mi sforzi, non riesco a imparare l’inglese & No matter how hard I try, I cannot learn English & 6-point Likert & \texttt{cant\_learn\_english\textsuperscript{\dag}} \\ \midrule
* Quando un esercizio è troppo difficile, lo salto oppure faccio solo le parti più facili. & When an exercise is too difficult, I skip it or just do the easier parts. & 6-point Likert & \texttt{skip\_hard\_exercises\textsuperscript{\dag}} \\ \midrule
* Quando un argomento è troppo difficile, lo salto e non provo a impararlo. & When a topic is too difficult, I skip it and do not try to learn it. & 6-point Likert & \texttt{skip\_hard\_topics\textsuperscript{\dag}} \\ \midrule
Sono portato per le lingue. & I am talented for languages. & 6-point Likert & \texttt{good\_at\_languages} \\
\midrule
\multicolumn{4}{c}{\textbf{ARCS - Homework}} \\
\midrule
I compiti di inglese sono noiosi. & English homework is boring. & 6-point Likert & \texttt{homework\_boring\textsuperscript{\dag}} \\ \midrule
I compiti di inglese sono utili per migliorare il mio livello di inglese. & English homework is useful for improving my level of English. & 6-point Likert & \texttt{homework\_improves\_english} \\ \midrule
I compiti di inglese sono utili per ottenere conoscenze che mi serviranno in futuro. & English homework is useful for gaining knowledge that will be useful in the future. & 6-point Likert & \texttt{homework\_future\_knowledge} \\ \midrule
Ho le capacità personali per portare a termine correttamente i compiti di inglese tutte le volte. & I have the personal abilities to properly complete English homework every time. & 6-point Likert & \texttt{ability\_finish\_homework} \\ \midrule
Avrei bisogno di più supporto e risorse per riuscire a fare i compiti di inglese. & I would need more support and resources to be able to do English homework. & 6-point Likert & \texttt{need\_support\_homework\textsuperscript{\dag}} \\ \midrule
Sono soddisfatto di quello che imparo facendo i compiti di inglese. & I am satisfied with what I learn from doing English homework. & 6-point Likert & \texttt{satisfied\_with\_homework} \\
\midrule
\multicolumn{4}{c}{\textbf{ARCS - Lectures and Contents}} \\
\midrule
I contenuti del programma di inglese non mi interessano. & The contents of the English program do not interest me. & 6-point Likert & \texttt{uninterested\_in\_content\textsuperscript{\dag}} \\ \midrule
Le lezioni di inglese sono utili a migliorare il mio livello di inglese. & English lessons are useful in improving my level of English. & 6-point Likert & \texttt{lessons\_improved\_english} \\ \midrule
Le lezioni di inglese mi forniscono conoscenze utili per il mio futuro. & English lessons provide me with useful knowledge for my future. & 6-point Likert & \texttt{lessons\_future\_knowledge} \\ \midrule
Ho le capacità personali per stare al passo col programma di inglese. & I have the personal abilities to keep up with the English program. & 6-point Likert & \texttt{ability\_keep\_up} \\ \midrule
Avrei bisogno di più supporto e risorse per riuscire a stare al passo col programma di inglese. & I would need more support and resources to be able to keep up with the English program. & 6-point Likert & \texttt{need\_support\_to\_keep\_up\textsuperscript{\dag}} \\ \midrule
Complessivamente, sono soddisfatto di quello che imparo a scuola in inglese. & Overall, I am satisfied with what I learn in English at school. & 6-point Likert & \texttt{overall\_satisfaction} \\ \midrule
\end{longtable}

\end{scriptsize}

\begin{scriptsize}
\begin{longtable}{@{}p{0.25\linewidth}p{0.25\linewidth}p{0.2\linewidth}p{0.2\linewidth}@{}}
\caption{Final Questionnaire}
\label{tab:app_final_q}
\\
\toprule
\textbf{Original Question} & \textbf{Translation} & \textbf{Type of Answer} & \textbf{Dataset Label} \\
\midrule
\endfirsthead
\multicolumn{4}{c}%
{\tablename\ \thetable\ -- \textit{Continued from previous page}} \\
\toprule
\textbf{Question (Italian)} & \textbf{Question (English Translation)} & \textbf{Type of Answer} & \textbf{Label} \\
\midrule
\endhead
\midrule \multicolumn{4}{r}{\textit{Continued on next page}} \\ \midrule
\endfoot
\bottomrule
\endlastfoot

Da quale dispositivo hai eseguito l'accesso alla piattaforma per i compiti per casa? & From which device did you access the platform for homework? & Mobile, Laptop, Both & \texttt{device\_used} \\
\midrule
\multicolumn{4}{c}{\textbf{General Questions}: Questions marked with a `*' are standardized SESQ.} \\
\midrule
Nelle ultime 8 settimane, ho fatto fatica a stare al passo col programma di inglese & In the last 8 weeks, I have struggled to keep up with the English program & 6-point Likert & \texttt{struggled\_with\_english\textsuperscript{\dag}} \\ \midrule
Voglio raggiungere un buon livello di inglese nei prossimi anni (approssimativamente un C1) & I want to reach a good level of English in the coming years (approximately a C1) & 6-point Likert & \texttt{goal\_c1\_english} \\ \midrule
Penso di avere le capacità per raggiungere un buon livello di inglese (approssimativamente un C1) nel corso dei prossimi anni & I think I have the ability to reach a good level of English (approximately a C1) over the coming years & 6-point Likert & \texttt{confidence\_in\_ability} \\ \midrule
* In inglese, indipendentemente da quanto un argomento è difficile, sono sicuro di poterlo capire & In English, no matter how difficult a topic is, I am confident I can understand it & 6-point Likert & \texttt{confident\_in\_english} \\ \midrule
* Non sono sicuro di poter imparare gli argomenti più difficili del programma di inglese & I am not sure I can learn the more difficult topics of the English program & 6-point Likert & \texttt{doubt\_difficult\_topics\textsuperscript{\dag}} \\ \midrule
* Sono sicuro di poter andare bene in verifiche, interrogazioni e/o test di inglese & I am confident that I can do well in quizzes, oral exams, and/or English tests & 6-point Likert & \texttt{confident\_in\_tests} \\ \midrule
* Per quanto mi sforzi, non riesco a imparare l’inglese & No matter how hard I try, I cannot learn English & 6-point Likert & \texttt{cant\_learn\_english\textsuperscript{\dag}} \\ \midrule
* Quando un esercizio è troppo difficile, lo salto oppure faccio solo le parti più facili & When an exercise is too difficult, I skip it or only do the easier parts & 6-point Likert & \texttt{skip\_hard\_exercises\textsuperscript{\dag}} \\ \midrule
* Quando un argomento è troppo difficile, lo salto e non provo a impararlo & When a topic is too difficult, I skip it and do not try to learn it & 6-point Likert & \texttt{skip\_hard\_topics\textsuperscript{\dag}} \\ \midrule
Sono portato per le lingue & I am talented at languages & 6-point Likert & \texttt{good\_at\_languages} \\ \midrule
\midrule
\multicolumn{4}{c}{\textbf{ARCS Homework}} \\
\midrule
Nelle ultime 8 settimane, i compiti di inglese sono stati noiosi & In the last 8 weeks, the English homework has been boring & 6-point Likert & \texttt{homework\_boring\textsuperscript{\dag}} \\ \midrule
Nelle ultime 8 settimane, i compiti di inglese sono stati utili per migliorare il mio livello di inglese & In the last 8 weeks, the English homework has been useful for improving my level of English & 6-point Likert & \texttt{homework\_improves\_english} \\ \midrule
Nelle ultime 8 settimane, i compiti di inglese sono utili per ottenere conoscenze che mi serviranno in futuro & In the last 8 weeks, the English homework has been useful for gaining knowledge that will be useful in the future & 6-point Likert & \texttt{homework\_future\_knowledge} \\ \midrule
Nelle ultime 8 settimane, ho sentito di avere le capacità personali per portare a termine correttamente i compiti di inglese tutte le volte & In the last 8 weeks, I have felt that I personally had the abilities to properly complete the English homework every time & 6-point Likert & \texttt{ability\_finish\_homework} \\ \midrule
Nelle ultime 8 settimane, avrei avuto bisogno di più supporto e risorse per fare i compiti di inglese & In the last 8 weeks, I would have needed more support and resources to do the English homework & 6-point Likert & \texttt{need\_support\_homework\textsuperscript{\dag}} \\ \midrule
Sono soddisfatto di quello che ho imparato facendo i compiti di inglese nelle ultime 8 settimane & I am satisfied with what I have learned from doing the English homework in the last 8 weeks & 6-point Likert & \texttt{satisfied\_with\_homework} \\ \midrule
\midrule
\multicolumn{4}{c}{\textbf{ARCS Lectures and Contents}} \\
\midrule
Nelle ultime 8 settimane, i contenuti del programma di inglese non mi interessavano & In the last 8 weeks, the contents of the English program did not interest me & 6-point Likert & \texttt{uninterested\_in\_content\textsuperscript{\dag}} \\ \midrule
Nelle ultime 8 settimane, le lezioni di inglese sono state utili a migliorare il mio livello di inglese & In the last 8 weeks, the English lessons have been useful in improving my level of English & 6-point Likert & \texttt{lessons\_improved\_english} \\ \midrule
Nelle ultime 8 settimane, le lezioni di inglese mi hanno fornito conoscenze utili per il mio futuro & In the last 8 weeks, the English lessons have provided me with useful knowledge for my future & 6-point Likert & \texttt{lessons\_future\_knowledge} \\ \midrule
Nelle ultime 8 settimane, ho sentito di avere le capacità personali per stare al passo col programma di inglese & In the last 8 weeks, I have felt that I personally had the abilities to keep up with the English program & 6-point Likert & \texttt{ability\_keep\_up} \\ \midrule
Nelle ultime 8 settimane, avrei avuto bisogno di più supporto e risorse per riuscire a stare al passo col programma di inglese & In the last 8 weeks, I would have needed more support and resources to keep up with the English program & 6-point Likert & \texttt{need\_support\_to\_keep\_up\textsuperscript{\dag}} \\ \midrule
Complessivamente, sono soddisfatto di quello che ho imparato in inglese a scuola & Overall, I am satisfied with what I have learned in English at school & 6-point Likert & \texttt{overall\_satisfaction} \\ \midrule
\midrule
\multicolumn{4}{c}{\textbf{Questions for Treatment Group Only}} \\
\midrule
Trovi che il tutor ti sia stato d’aiuto nello svolgere i compiti per casa? & Did you find the tutor helpful in doing homework? & Yes; No; No Access & \texttt{tutor\_helped\_homework} \\ \midrule
Trovi che il tutor ti abbia aiutato a migliorare l’inglese ad un livello pratico? & Did you find that the tutor helped you improve English to a practical level? & Yes; No; No Access & \texttt{tutor\_improved\_english} \\ \midrule
Trovi che avere accesso al tutor ti abbia aiutato a stare al passo col programma di inglese? & Did having access to the tutor help you keep up with the English program? & Yes; No; No Access & \texttt{tutor\_helped\_keep\_up} \\ \midrule
Vorresti continuare ad avere accesso al tutor in futuro? & Would you like to continue having access to the tutor in the future? & Yes, for all exercises; Yes, for some exercises; No; No access & \texttt{continue\_with\_tutor} \\ \midrule
Quali aspetti del tutor hai trovato utili? & Which aspects of the tutor did you find useful? & Options: Personalized explanations; Feedbacks and Corrections on My Answers; Guidance through the Exercise Step by Step & \texttt{useful\_tutor\_aspects} \\ \midrule
Spesso, il tutor mandava messaggi troppo lunghi anche quando non era necessario & Often, the tutor sent messages that were too long even when not necessary & 6-point Likert & \texttt{tutor\_long\_messages} \\ \midrule
Spesso, il tutor diceva cose non vere oppure mi diceva che sbagliavo anche quando la mia risposta era corretta & Often, the tutor said things that were not true or told me I was wrong even when my answer was correct & 6-point Likert & \texttt{tutor\_wrong\_feedback} \\
\midrule
\multicolumn{4}{c}{\textbf{Final Comments}} \\
\midrule
Hai qualche altro commento sull'esperienza in generale? & Do you have any other comments on the general experience? & Free text response & \texttt{general\_comments} \\ \midrule

\end{longtable}
\end{scriptsize}

\begin{scriptsize}
\begin{longtable}{@{}p{0.25\linewidth}p{0.25\linewidth}p{0.2\linewidth}p{0.2\linewidth}@{}}
\caption{Weekly Questionnaire}
\label{tab:app_weekly_q}
\\
\toprule
\textbf{Question} & \textbf{Translation (English)} & \textbf{Type of Answer} & \textbf{Label} \\
\endfirsthead
\multicolumn{4}{c}%
{\tablename\ \thetable\ -- \textit{Continued from previous page}} \\
\toprule
\textbf{Question} & \textbf{Translation (English)} & \textbf{Type of Answer} & \textbf{Label} \\
\midrule
\endhead
\midrule \multicolumn{4}{r}{\textit{Continued on next page}} \\ \midrule
\endfoot
\bottomrule
\endlastfoot

\midrule
\multicolumn{4}{c}{\textbf{Week-level Questions}}\\
\midrule
Hai fatto, o provato a fare, almeno uno degli esercizi assegnati questa settimana? (rispondi sinceramente, non condivideremo la risposta con la tua insegnante) & Have you done, or tried to do, at least one of the exercises assigned this week? (answer honestly, we will not share the response with your teacher) & Yes; No & \texttt{completion} \\ \midrule
I compiti per casa di questa settimana erano interessanti e/o stimolanti (rispetto ai compiti per casa prima dell'esperimento). & This week's homework was interesting and/or stimulating (compared to homework before the experiment). & 6-point Likert & \texttt{\TI} \\ \midrule
I compiti per casa di questa settimana sono stati utili a migliorare il mio inglese ad un livello pratico. & This week's homework has been useful in improving my English to a practical level. & 6-point Likert & \texttt{\TU} \\
\midrule
\multicolumn{4}{c}{\textbf{Exercise-Specific Questions} - These questions are repeated for each exercise.}\\
\midrule
Questo esercizio è stato utile a migliorare la mia comprensione e la mia conoscenza dell'argomento trattato. & This exercise was useful in improving my understanding and knowledge of the topic covered. & 6-point Likert & \texttt{\EC} \\ \midrule
Avevo a disposizione supporto e risorse a sufficienza per risolvere adeguatamente questo esercizio (spiegazioni, materiali, ...) & I had enough support and resources available to adequately solve this exercise (explanations, materials, ...). & 6-point Likert & \texttt{\EI} \\ \midrule

(\textbf{Treatment-Group Only}) Quali aspetti hai trovato utili? (seleziona tutte le opzioni rilevanti) & (\textbf{Treatment-Group Only}) Which aspects did you find useful? (select all relevant options) & Options: Personalized explanations; Feedbacks and Corrections on My Answers; Guidance through the Exercise Step by Step & \texttt{useful\_aspects\_1} \\
\midrule
\multicolumn{4}{c}{\textbf{Anomalies Feedback}}\\
\midrule
Il sistema ha avuto dei comportamenti anomali? (se sì, puoi descriverli brevemente?) & Did the system exhibit any abnormal behaviors? (if yes, can you briefly describe them?) & Open-ended & \texttt{anomalies} \\
\end{longtable}
\end{scriptsize}
\subsection{Regression Tables}

\begin{table}[h]
\centering

\begin{tabular}{lcc}
\hline
\textbf{Variable} & \textbf{Coefficient} & \textbf{Std. Error} \\ \hline
Intercept    & 3.5828  & 0.743  \\
Treatment    & -0.4456 & 1.027  \\
3rd Year     & 4.0815  & 0.852  \\
Words Typed  & 0.001   & 0.001  \\ \hline
\end{tabular}
\caption{OLS Regression Results: Learning Gains Mediated by Student Engagement}
\label{tab:eng}
\end{table}

\section{Prompts}
\label{app:prompt}
\subsection{Tutoring Prompt}
The following prompt was the main prompt given to the tutor as the system prompt. We replaced assignment purpose, description and example with the content provided from the teacher. Note that the Common European Framework of Reference for Languages was mistakenly referred to as ``Cambridge Framework'' during the execution of the study.

\begin{scriptsize}
\texttt{\\
We are helping students learn english as a second language. \\
\\
We give you an exercise as a starting point. Act as a tutor and drive the student through the same concepts, testing the understanding step by step. It is not necessary to replicate to the example exercise, as long as you cover the same concepts. Follow the tutoring strategy we provide. \\
\\
Start with a brief explanation of the concept. \\
Provide at least 10 questions, one by one. \\
Do not move on until the student gives the correct answer. \\
If necessary, provide explanations and feedback. \\
Never give the answer to the question. \\
Do not give the answer to the question as a part of the explanation. \\
Point out all grammar and spelling mistakes. \\
Keep a B2 level of English according to the Cambridge framework. \\
\\
Once all questions are solved, ask the student if they wish to practice more. If they don't, output <COMPLETE> \\
\\
EXERCISE PURPOSE: \\
{+++ purpose +++} \\
\\
EXERCISE DESCRIPTION \\
{+++ description +++} \\
\\
EXERCISE EXAMPLE (NOT VISIBLE TO THE STUDENT) \\
{+++ example +++} \\
\\
TUTORING STRATEGY \\
{+++ strategy +++} \\
\\}

\end{scriptsize}
\subsection{Strategy Generation Prompt}
The following prompt was used to generate the tutoring strategy.

\begin{scriptsize}
\texttt{\\
We are helping students learn english as a second language.\\
\\
We give you an exercise as a starting point.\\
Provide a concise, step-by-step strategy for a short dialog-based tutoring session led by ChatGPT with a student covering the same concepts.\\
The tutoring session is text-based and led through a chat interface.\\
Describe the strategy with a maximum of six sentences.\\
\\
Keep a B2 level of english according to the Cambridge framework.\\
+++ assignment.purpose +++\\
+++ assignment.description +++\\
+++ assignment.example +++\\
}

\end{scriptsize}

\section{Homework Content}
We report the content of homework for each class and each week.
\begin{scriptsize}
\begin{longtable}{p{0.10\textwidth} p{0.43\textwidth} p{0.43\textwidth}}

\toprule\\
       & \textbf{3rd Year}                                                                                                           & \textbf{5th Year}                                                                                                        \\ \midrule
\textbf{Unit 0} & -                                                                                                                 & \textbf{(Only class 5A) WWI, open questions}: Students answer questions about\label{tab:homework} WWI and British responses in 8-10 lines. \\ \midrule
\textbf{Unit 1} & \textbf{Conditionals, sentence correction}: Students are asked to fix mistakes in the given sentences.            & \textbf{Poem commentary, analysis, and comparison}: Students analyze and compare “The Soldier” by R. Brooke and “Dulce et Decorum est” by W. Owen. \\
                & \textbf{Conditionals, sentence completion}: Students complete sentences using the correct tense of the verb in brackets. & \\
                & \textbf{Conditionals, sentence transformation}: Students complete the second sentence to have the same meaning as the first using the given word. & \\ \midrule
\textbf{Unit 2} & \textbf{I wish/If only and mixed conditionals, either/or questions}: Students choose the correct verb form from two options for sentences using "I wish," "If only," and mixed conditionals. & \textbf{Key events of the 20th century, open questions}: Students answer questions about 1967, WWII, and Margaret Thatcher in 4-6 lines. \\
                & \textbf{I wish/If only and mixed conditionals, sentence completion}: Students complete sentences using the correct tense of the verb in brackets. & \\ \midrule
\textbf{Unit 3} & \textbf{Essay on contemporary social issues}: Students write a 200-220 word argumentative essay on a contemporary social issue. & \textbf{The Roaring 20s, open questions}: Students answer questions about the economic, political, and social changes in the US during the 20s, including “The Great Gatsby.” \\ \midrule
\textbf{Unit 4} & \textbf{Passive voice, sentence transformation}: Students transform active sentences to passive form while keeping the same tense. & \textbf{Mid-century America, open questions}: Students answer questions about the Cold War, the 60s, cultural revolution, and the crisis of the 70s in the US. \\
                & \textbf{Passive voice, sentence completion}: Students complete sentences using the correct tense of the verb in brackets, focusing on passive forms. & \\
                & \textbf{Passive vs. active voice, either/or questions}: Students choose the correct active or passive form and tense from two options given. & \\ \midrule
\textbf{Unit 5} & \textbf{Causative verbs (have/get something done), sentence completion}: Students complete sentences using the correct tense of the verb in brackets, focusing on the causative form. & \textbf{Human rights movements, Gandhi, open questions}: Students answer questions about Gandhi's life, achievements, and influence in 4-6 lines. \\
                & \textbf{Passive voice with double object, sentence transformation}: Students transform active sentences with two objects to passive form, giving both options. & \\ \midrule
\textbf{Unit 6} & \textbf{Reported speech statements, sentence transformation}: Students transform direct statements into reported speech, using the correct tense based on the introductory verb. & \textbf{(Only class 5B) Important women in history, open questions}: Students answer questions about Queen Victoria, Emmeline Pankhurst, and Rosa Parks in 4-6 lines. \\
                & \textbf{Reported speech questions, sentence transformation}: Students transform direct questions into reported speech, using the correct tense based on the introductory verb. & \\
                & \textbf{Reported speech correction, sentence correction}: Students correct the mistakes in sentences related to reported speech using different reporting verbs. & \\ \bottomrule
\end{longtable}
\end{scriptsize}
\end{document}